\documentclass{svjour2}                    % onecolumn
\smartqed  % flush right qed marks, e.g. at end of proof
\usepackage{graphicx}
\begin{document}

\title{Quantum Computing's Classical Problem, Classical Computing's
  Quantum Problem~\thanks{This work was supported by the Japan Society
    for the Promotion of Science (JSPS) through its ``Funding Program
    for World-Leading Innovative R\&D on Science and Technology (FIRST
    Program).''
% Grants or other notes
}
}
% \subtitle{Classical Computing's Quantum Problem}


\author{Rodney Van Meter
}


\institute{R. Van Meter \at
  Keio University \\
              5322 Endo \\
              Fujisawa, Kanagawa 252-0882 JAPAN \\
              Tel.: +81-466-49-1100\\
              \email{rdv@sfc.wide.ad.jp}           %  \\
}

\date{Received: date / Accepted: date}

\maketitle

\begin{abstract}
  Tasked with the challenge to build better and better computers,
  quantum computing and classical computing face the same conundrum:
  the success of classical computing systems.  Small quantum computing
  systems have been demonstrated, and intermediate-scale systems are
  on the horizon, capable of calculating numeric results or simulating
  physical systems far beyond what humans can do by hand.  However, to
  be commercially viable, they must surpass what our wildly
  successful, highly advanced classical computers can already do.  At
  the same time, those classical computers continue to advance, but
  those advances are now constrained by thermodynamics, and will soon
  be limited by the discrete nature of atomic matter and ultimately
  quantum effects.  Technological advances benefit both quantum and
  classical machinery, altering the competitive landscape.  Can we
  build quantum computing systems that out-compute classical systems
  capable of some $10^{30}$ logic gates per month?  This article will
  discuss the interplay in these competing and cooperating
  technological trends.

\end{abstract}

\section{Introduction}
\label{sec:intro}

Imagine, for a moment, that classical computers did not exist, and
that quantum computers were being developed in a technological Garden
of Eden, innocent of the taint of electronic or mechanical
computation.  What a fantastic future would await!  On the horizon,
machines that could solve specialized mathematical problems, search
through large spaces of possible problem solutions without iterating
over the entire space, and calculate numeric values describing (other)
quantum
systems~\cite{bacon10:_recent_progress,feynman:_simul_physic_comput,grover96,hallgren2007ptq,harrow:lineqs,jordan12:qa-qft,magniez2005quantum,mosca2008quantum,shor:factor,whitfield2011simulation}.
Surely such an enticing set of capabilities would be enough to lure us
out of our pre-technological paradise.

Instead, advances toward quantum computing machines are taking place
in a world already filled with fantastic classical computing machines
of almost unimaginable power.  It has become trite to make comparisons
of the computational power of a smart phone to that used to guide an
Apollo moon mission, but it is no less true for that.  The
computational power of a modern-day supercomputer is nothing short of
astonishing, and sets an extraordinarily high bar against which the
utility of prospective commercial quantum computing systems will be
measured.

In this article, we first discuss the market challenges faced by
attempts to build a quantum computer, by comparing to the technical
capabilities of existing and prospective supercomputers.  After this
somewhat pessimistic note, we turn to the challenges that classical
computing faces in its continuing technological evolution.  We end by
presenting some historical perspective on these two issues, and reason
for optimism on both fronts.

\section{Quantum Computing's Classical Problem}

In order to be economically attractive, a quantum computer must be
able to compute some output that cannot be calculated using classical
computers and supercomputers.  Classical supercomputers have two main
advantages over quantum computers: computational capability measured
in logic gates or floating point operations per second (FLOPS), and
memory capacity, which together determine the type and scale of
problems which can be successfully attacked.  In the last quarter of a
century, large-scale computations have come to be dominated by
parallel and distributed computing systems, on which we will focus
here~\cite{anderson2004bsp,asanovic2006landscape,coulouris:distributed-systems-4ed,fox1994parallel,hennessy-patterson:arch-quant4ed}.

In accordance with the Gustafson-Barsis
Law~\cite{gustafson88:_reeval_amdahl}, as the size of a supercomputer
grows, the size of problems that it can tackle grows.  Japan's K (Kei)
supercomputer is used for climate research, medical research, and
computational chemistry.  Weather modeling, protein folding, seismic
simulations, fluid flow, galactic evolution simulations, and
high-energy physics calculations are just a few of the common
applications of such large-scale systems.  What applications will a
quantum computer have that go beyond such capabilities?  To answer
this question, let us first probe more deeply what such classical
computing power means.

As an example of current capabilities, consider K, which from mid-2011
to mid-2012 was the most powerful system in the world, reaching a
speed of over 10 petaFLOPS on the LINPACK linear algebra benchmark
using 705,024 processing cores in 88,128 processors.

Each SPARC64 VIIIfx processor, fabricated using Fujitsu's 45nm
process, consists of 760 million transistors.  Thus, the processing
elements and on-chip cache memory, interconnects, etc. across the
entire system comprise approximately $6.7\times 10^{13}$ transistors.
The main memory, allocating a single transistor per DRAM cell, is some
$1.1\times 10^{16}$ transistors, nearly two hundred times as large.

Considering only the processing elements, the rate of floating point
operations alone is some $10^{16}$ 64-bit operations per second that
can be deployed on solving a single classical problem.  An alternative
way of considering performance is the rate of single-bit classical
gate operations.  If we assume that an average of $\sim 10\%$ of the
transistors in the processor are switching in any given clock cycle,
we have $6.7\times 10^{12}$ switches per clock cycle.  With a clock
speed of 2.0GHz, we have $1.3\times 10^{22}$ switches, or gates, per
second.  One large-scale computation may take as much as a month on
such a system.  A month is approximately 2.5 million seconds, giving
us a total computation capacity of approximately $3\times 10^{28}$
logic gates per month.

The quest to build economically attractive quantum computers must
focus on workloads consisting of problems that cannot be solved using
such enormous, powerful systems.  If we target deploying such a
quantum computer in the year 2020, we must strive to exceed the
classical systems that will be available then.  Supercomputing system
engineers are promising a system one hundred times as powerful by the
year 2020, a capacity of more than $10^{30}$ logic gates or $10^{24}$
FLOPs per month.

Shor's algorithm for factoring large numbers, the most famous quantum
algorithm to date, occupies something of a sweet spot in problem
difficulty: although execution time does not grow truly exponentially
with problem size (and is not an NP-complete problem), it does grow
superpolynomially.  Moreover, factoring is a problem for which
approximate solutions simply will not do: a number is never
``approximately'' an integer factor of another integer.  Only exact,
and very difficult to calculate, solutions are acceptable.

In constrast, many other computationally difficult problems, including
most of the applications of supercomputers, admit practically useful
approximate answers that can be calculated using a variety of
heuristic algorithms.  In many cases, especially simulations, the
problems in question have no exact answer that can be checked (and
hence are also not NP-complete problems), but increasing expenditure
of computational effort results in a closer match to some objective
reality that can be measured.  This results in improvements of the
accuracy of weather simulations, better airplane designs, and more
digits of confirmation of the accuracy of physical theories.

This, then, is the crux of quantum computing's classical problem:
classical systems are already extraordinarily successful at generating
results with enormous impact on science and technology, and through
them on society as a whole.  The search for useful and interesting
quantum algorithms then becomes a hunt for problems (a) that do not
admit approximate solutions and for which finding exact solutions
becomes hard quickly as the problem size grows, or (b) whose required
classical resources (logic gates, FLOPs or memory) grow more rapidly
as the problem size grows than even supercomputers can handle, {\em
  but} can be solved in a quantum computer that is many orders of
magnitude smaller in capacity and likely also far slower at executing
logical gates.  Shor fits the first category, and simulation of
quantum systems the second.  A few other algorithms have been
developed that likewise appear to fit in this sweet spot, and detailed
analysis of their needs has become a community priority.

In general, the ideal problem is one in which a classical solution
grows exponentially in the number of operations (computational volume)
as the problem size grows, but a quantum solution is polynomial.
However, the common language of computer science theorists, the
$O(\cdot)$ notation, can hide large constant factors, and even
seemingly innocent polynomials can grow quickly in practical terms.

Motivated in part by results that indicate that systems powerful
enough to run Shor's algorithm will be very large and disappointingly
slow~\cite{devitt09:arch,PhysRevX.2.031007,van-meter10:dist_arch_ijqi,whitney:isca09},
the U.S. government agency IARPA is funding a program to examine in
detail the possibilities of executing several other quantum
algorithms:
\begin{enumerate}
\item Ambainis {\em et al.}'s Boolean formula evaluation
  algorithm~\cite{ambainis2007any}.
\item Childs {\em et al.}'s binary welded tree algorithm, executed
  using a quantum random walk, with a tree height of
  300~\cite{childs2003eas}.
\item Hallgren's solution for the class number problem, for 124
  decimal digits~\cite{hallgren2007ptq}.
\item Linear systems of equations, with an array dimension of $3\times
  10^8$, using an adapted form of Harrow {\em et al.}'s algorithm for
  sparse matrices~\cite{harrow:lineqs}.
\item Magniez {\em et al.}'s triangle-finding
  algorithm~\cite{magniez2005quantum}.
\item The unique shortest vector (USV) problem, for a problem
  dimension of 50~\cite{regev2002quantum}.
\item Feynman's original proposal for quantum computers discussed
  simulating quantum systems~\cite{feynman:_simul_physic_comput};
  Whitfield {\em et al.} have developed a specific algorithm that can
  be used to find molecular ground-state
  energy~\cite{whitfield2011simulation}.  This algorithm is proposed
  to be used for an iron-sulphur molecular complex with 208 basis
  functions, to find the result to 9 bits of accuracy.
\end{enumerate}

Each of these was given a specific problem instance, intended to
represent a post-classical result: the size and complexity were set at
a level which is considered to be impractical for classical systems.
In some cases, this involved modification of parts of the algorithm to
include preparation of difficult-to-create input states.  These
algorithms are then matched with a physical technology and a quantum
error correction mechanism.

Early results from the four teams working on this project suggest that
all of these algorithms will be computationally challenging for
quantum computers.  However, the program (along with the work of other
groups) is already paying dividends in the form of new approaches to
single-gate decomposition that promise to reduce execution times by
orders of
magnitude~\cite{bocharov2012depth,pham12:_optim_solov_kitaev,selinger12:_effic_cliff_T}.
However, further advances in architecture, compilation, and especially
error correction, as well as the underlying physical technologies, are
necessary to bring most of these applications within
reach~\cite{van-meter06:thesis,van-meter13:_blueprint}.

Were it not for a critical technical factor, perhaps the story of
quantum computing would end there, with a handful of specialized
algorithms and a very high hurdle to clear in order to reach economic
viability.  However, just as quantum computing has a classical
problem, classical computing has a quantum problem.

\section{Classical Computing's Quantum Problem}

This problem could be more completely described as classical
computing's quantum and atomic problem, and can be summarized
succinctly: semiconductor technology is a victim of its own
success~\cite{ITRS2012}.  After conquering the basic theoretical and
fabrication problems in the period from the late 1940s through the
early 1960s, integrated circuits entered a Golden Age of tremendous
growth in density, known as Moore's Law~\cite{moore65}.  During this
period, economic incentives and technical genius combined to raise the
number of transistors that could be fabricated in a given area.
Capacities doubled approximately every two years, and operating speeds
likewise increased, reaching billion-transistor chips running at
gigahertz speeds by the early 2000s.

In the mid-2000s, we entered what I call the Late Moore's Law Period.
The rate of density improvement declined to doubling every 36 months
or so, instead of every 24 months.  More importantly, clock speeds
stalled almost entirely.  Typical CPU speeds have remained near 2GHz
since then, although improvements to cache and other peripheral
functions have brought modest improvements in the average amount of
work completed in each clock cycle.  Instead, the focus has shifted to
improving functionality on important, computationally-intensive tasks
such as graphics.  This is achieved by increasing the parallelism in
the system at both the micro and macro levels, adding both
general-purpose cores and graphics processing unit (GPU) cores.

What is driving this decline, and is it temporary or permanent?  The
end of Moore's Law has been predicted repeatedly since its
inception~\cite{forbes2003guest,kish2002ems,tuomi2002lives}.  Reasons
cited include concerns about the technical difficulty of
short-wavelength photolithography, about the economic viability of
increasingly expensive fab lines, and the difficulty of making
defect-free devices of such densities.  Such concerns are predicated
on engineering and economic matters, which are of course critical, but
are less absolute pronouncements than those based on fundamental
physical characteristics.  In the latter category, two issues loom
large: the atomic (and ultimately quantum) nature of matter, and
Landauer's limit on the thermodynamic cost of irreversible (Boolean)
logic~\cite{ITRS2012,landauer61:_irreversibility,meindl01:_terascale-si}.

Although I do not subscribe to a fully Kurzweilian world view, it is
true that our technology for computing has produced a generally
downward trend in memory and logic element size, extending back to the
beginning of computation.  Abacuses, developed three to four thousand
years ago, must consist of elements large enough for human fingers to
manipulate while storing a few bits of data.  The Inca quipu, dating
back some 500 years, stores information in the type and position of
knots in a string, necessarily at the human scale.  These two
innovations are among the first storage technologies for numeric data.

By the mid-seventeenth century, devices with at least some support for
mechanical computation emerged: the slide rule, and Blaise Pascal's
pascaline, a slow and error-prone mechanical adder.  In the early
nineteenth century, the Jacquard loom used a precursor to punch cards
to control the pattern woven into a textile.  Charles Babbage used
sets of mechanical rotors as data registers in his Difference Engine,
which was capable of calculating values for a chosen
polynomial~\cite{swade02:_difference_engine}.  Babbage's projects were
not completed in his lifetime, and although he is now revered his work
languished in some obscurity when modern computing began developing in
the early- to mid-twentieth century.

Up to this point, little if any trend in the size of storage or
computational elements is discernible, as all technologies were
limited by the ability of human hands to create and manipulate them.
Punch cards and paper tape shrank the volume of an individual bit, but
even replacing relays with vacuum tubes, as happened in the 1940s with
the introduction of ENIAC (the first electronic general-purpose
digital computer), improved the speed but not so much the size of
logic elements.  The transistor was invented in 1947.  Smaller and
cooler than a vacuum tube, it was the basis for a computer constructed
in Manchester in 1953.  But it was not until the development of the
integrated circuit in 1958 that density took a sharp upward turn, as
we developed tools capable of making devices for us using
non-mechanical means~\cite{riordan97:_cryst_fire}.  By 1965,
progress was steady enough that Gordon Moore could analyze the
economic imperatives behind chip manufacture and formulate his famous
law, which ultimately moved from passive analytic tool to business
dictum~\cite{moore65}.

In 1971, the world's first microprocessor, the Intel 4004, was
introduced, consisting of some 2,300 transistors fabricated in a
$10\mu{}m$ process.  This represented a thousand-fold increase in chip
capacity in thirteen years.  It would take eighteen years to reach a
million transistors in a logic chip (the Intel i486, in 1989), but
less than sixteen more years to reach a billion transistors in a chip,
with Intel shipping a 1.7 billion transistor microprocessor in 2005,
fabricated in a 90nm process.

As of late 2012, chips with a minimum feature size of 22nm are in
production~\footnote{Some care must be taken in comparing the exact
  feature sizes, as memory and logic chips are sometimes described
  using different terminology varying by a factor of two or so, and
  the actual feature size on chip may differ from the fabrication
  process, due to factors in the lithography and etching.}.
Engineering difficulties multiply as we get down toward 10nm and even
below~\cite{ieong04:_silic_devic_scalin_sub_regim,ITRS2012,lundstrom2003ape},
but we are not yet at fundamental limits.  The International Technical
Roadmap for Semiconductors currently projects that the minimum size of
a structure in a chip will reach 5.8nm in 2026, providing a
challenging but in some ways reassuring path over the next fourteen
years.

With a path that leads forward into the mists of time, in terms of
technology generations, why worry?  Won't we continue pressing forward
for the indefinite future?  After all, we have cleared every hurdle to
date.

The problem is that our problems are becoming increasingly
fundamental.  The distances within a transistor can be measured in
atoms; that 22nm is only about 40 times the size of the silicon
crystal lattice cell of 0.54nm.  The 5.8nm of 2026 is only some eleven
times the lattice cell.  The exact practical limit has not yet been
determined, but it is clear that we don't know how to build
transistors with parts less than one atom thick!  Moore's Law, in its
current form governing increased density of transistors in a
two-dimensional layout, \emph{must} end within the next human
generation.

In addition to the limits to general dimensions, the atomic nature of
matter poses other problems.  The behavior of semiconductors depends
critically on small amounts of dopants added to the base material.
Early models of dopant activity could treat the dopants as a uniform
change to a region of the material, but we have now reached the point
where the effect of individual dopant atoms matters, and their
placement is critical but hard to control.

The actual quantum nature of current-carrying electrons has also begun
to matter, as they can tunnel not only through barriers (a desired
phenomenon but one we need to control) but also into and out of the
wires.  Resistance is also a problem at this scale.

The second problem, thermodynamics, has already manifested itself in
systems, and is the key reason that clock speeds for individual CPUs
have stalled at around 2GHz.  Landauer showed that erasing a bit, as
is necessary in any logic operation that is not bijective, results in
an unavoidable increase in entropy, manifesting itself as waste heat.
This is a fundamental fact of the physical implementation of logic,
independent of the medium in which it is built.  The exact value for
our current and near-future devices depends on factors such as the
thermal conductivity of bulk
silicon~\cite{ieong04:_silic_devic_scalin_sub_regim,ITRS2012,meindl01:_terascale-si}.
DeBenedictis has estimated that we can ultimately reach $10^{22}$
FLOPS or $2\times 10^{26}$ logic gates per second, or $2.5\times
10^{28}$ FLOPS and $5\times 10^{32}$ gates per month, before
Landauer's limit stops us~\cite{debenedictis05:_reversible}.  The 2020
system predicted above is within a factor of one hundred of this
limit, depending on the cost assigned to a floating point operation.
Thus, we can effectively see the end of the evolution of classical,
Boolean supercomputers coming.

\section{Discussion}

With such dire and seemingly fundamental problems, is there any reason
for optimism, for either economically viable quantum computers or
continuing advances in classical systems as Moore's Law peters out?
In fact, there is no reason to believe that progress in the overall
field of computing systems will stop.

Landauer's disciple Bennett, joined by Feynman, Toffoli, and Fredkin,
devised reversible logic schemes in the 1970s and
80s~\cite{bennett73:reversible,bennett88:_notes,feynman:lect-computation,fredkin82:_conserv_logic}.
Reversible logic results in no erasure of information, and therefore
no waste heat.  Reversible logic offers one path to continued
improvement, and both theorists and experimentalists are working on
making it
practical~\cite{athas:reversible,burignat12:_rev_review,frank99:_thesis,peres85:_reversible,shende03:_synth_reversible,vieri98:fully}.

(By interesting coincidence, Babbage's Difference Engine is logically
reversible: from any point in the evolution of the state of the
machine, it is possible to infer its state at any earlier point in
time.  It is, however, neither mechanically nor thermodynamically
reversible, and surely the thought of reversibility was not on
Babbage's mind when he developed the machine.)

Technologies with ambition to augment or replace standard CMOS
circuits abound in the labs; they are too numerous to cover
exhaustively~\cite{bourianoff03:_future_nanocomp,tseng2001net}.
Carbon nanotubes, numerous new types of transistors and ultimately
three-dimensional integration all promise to bring us new capabilities
within the same physical
constraints~\cite{beckman05:_bridg_dimen,dehon03:_array_based_nano,aaronson:thesis,topol2006three}.
One or more of these may succeed, and certainly there is no reason to
believe that our ingenuity in building \emph{systems} out of these
constrained \emph{technologies} has been exhausted.

The prospects for quantum computing systems continue to improve.  Our
understanding of how to develop algorithms has grown, compilation
techniques are improving, and new error correction mechanisms have
been developed.  The underlying physical technology steadily gets
better, in both memory lifetime and gate fidelity, and in technologies
such as superconducting systems has now reached the point where
moderate-scale demonstrations of quantum error correction can be
contemplated.  Importantly, work on the architecture of systems is
attracting increasing interest from the experimental
community~\cite{van-meter13:_blueprint}.  In early 2013, we seem to
stand on the cusp of an inflection point in experimental capabilities,
and the next few years likely will see a very competitive atmosphere
with important milestones achieved.

The ongoing evolution of classical technologies will benefit quantum
systems, especially once industrial processes can be applied.
Microcavity ring resonators, for example, depend on essentially
atomic-level surface smoothness, difficult to achieve in the lab but
potentially doable in an industry setting.

Working on quantum systems also tells us how to build better classical
systems in the light of quantum effects and thermodynamic
limitations~\cite{PhysRevX.2.031016}.  This is true at the physical
level as well as at the logical level, where work on reversible
circuits such as arithmetic, while often ostensibly focused on quantum
computing, applies to reversible classical as well.  More spinoff
benefits from research on quantum computing systems can be expected.

\section{Perspective}

We are approaching the 100th anniversary of the coining of the term
\emph{robot}~\cite{capek:RUR}, but Asimovian androids do not
(generally) roam the streets of Tokyo.  The field of robotics as a
whole, however, has contributed enormously to society's well being in
ways such as improved manufacturing and automated monitoring systems,
and to fields such as exploration of our solar system.  So I expect it
to be with quantum computing: development will take time, and the end
results and largest societal impact very likely will be nothing we can
anticipate today.

The invention (discovery?) of Shor's algorithm coincided with the
availability of a slew of experimental technologies on the verge of
single-quantum effects, if not actual digital, entangled operations.
This resulted in a surge in interest, and in funding.  Funding remains
primarily the domain of government agencies.  The total spent
worldwide on quantum computing research since 1995 is probably a
couple of billion dollars.  The annual R\&D budget of Intel alone is
some four times that amount.  When quantum system demonstrations
(including applications) reach a certain level of maturity, industrial
levels of investment will likely occur and the rate of progress will
accelerate.

A direct, frontal assault on the bastions of classical supercomputing
is unlikely to succeed.  Instead, as in \emph{The Innovator's
  Dilemma}, quantum computers may have to take over stealthily, by
emphasizing their strengths in areas of classical
weakness~\cite{christensen1997innovator}.  Quantum computers will not
compete to replace classical supercomputers directly, but instead will
open new avenues of computational and intellectual query.

Consistent funding and intelligent choice of problems to attack,
and a long-term view of the scale of systems that must be built
coupled with an impatience to solve problems and deploy systems, will
ultimately lead to success in solving both quantum computing's
classical problem and classical computing's quantum (and atomic)
problem.

\bibliographystyle{spphys}       % APS-like style for physics
% \bibliography{paper-reviews}   % name your BibTeX data base

\end{document}